\documentclass{epl}

\begin{document}

\title{Cross-correlation in financial dynamics}
\author{J. Shen and B. Zheng\thanks{corresponding author; email: zheng@zimp.zju.edu.cn}}

\institute{Zhejiang University, Zhejiang Institute of Modern
Physics, Hangzhou 310027, PRC}


\maketitle

\begin{abstract}
To investigate the universal structure of interactions in financial
dynamics, we analyze the cross-correlation matrix $\bf C$ of price
returns of the Chinese stock market, in comparison with those of the
American and Indian stock markets. As an important emerging market,
the Chinese market exhibits much stronger correlations than the
developed markets. In the Chinese market, the interactions between
the stocks in a same business sector are weak, while extra
interactions in unusual sectors are detected. Using a variation of
the two-factor model, we simulate the interactions in financial
markets.
\end{abstract}

In recent years, there has been a growing interest of physicists in
economic systems. Concepts and methods in physics have been applied
to the study of financial time series
\cite{man95,gop99,liu99,bou01,gab03,ren03,qiu06}. Different models
and theoretical approaches have been developed to describe the
features of the financial dynamics
\cite{cha97,lux99,sta99,con00,egu00,muz00,gia01,lou01,kra02,zhe04b,zhe04,ren06a}.
Statistical properties of price fluctuations and correlations
between different stocks are topics of interest, not only
scientifically for understanding the complex structure and dynamics
of the economy, but also practically for the asset allocation and
portfolio risk estimation \cite{far99,man00,bou03}. The probability
distributions of stock prices in different stock markets show a
universal nature and follow the "inverse cubic law"
\cite{lux96,ple99a,pan07}. However, the statistical properties of
correlations between different stocks seem less universal across
different stock markets \cite{pan07a}.

Unlike most traditional physical systems, where one derives
correlations between subunits from their interactions, the
underlying "interactions" for the stock markets are not known.
Pioneering studies at the phenomenological level analyze
cross-correlations between stocks by applying concepts and methods
of the random matrix theory (RMT), which was developed in the
context of complex quantum systems where the precise nature of the
interactions between subunits is not known \cite{meh91,guh98}. The
properties of the empirical correlation matrix $\bf C$ of price
returns are compared with those of a random matrix in which the
price movements are uncorrelated \cite{lal99,ple99}. This spectral
property-focused method was first applied to developed markets such
as the New York Stock Exchange (NYSE) in USA
\cite{lal99,ple99,ple02a,uts04}, and recently also to some emerging
markets, e.g. the National Stock Exchange (NSE) in India
\cite{pan07a}.

In general, the bulk of the eigenvalue spectrum of the correlation
matrix $\bf C$ of price returns shares universal properties with the
Gaussian orthogonal ensemble of random matrices, while the largest
eigenvalue of $\bf C$ which deviates significantly from the bulk
represents the influence of the entire market on all stocks. This is
true for both developed and emerging stock markets. For developed
markets, other large eigenvalues can be associated with the
conventionally identified business sectors, according to the
compositions of the eigenvectors. These sectors are stable in time,
in some cases, as long as 30 years \cite{gop01,ple02a}. But for
emerging markets such as the NSE in Indian, the number of large
eigenvalues are smaller than that of developed markets
\cite{pan07a}. Although there exist often much more correlations
between stocks in emerging markets in all, the less large
eigenvalues may reflect less correlations between stocks in a same
sector. It seems that strong interactions may emerge within groups
of stocks as a financial market evolves over time.

Although there have been many studies of correlated price movements
in stock markets, the researches on emerging markets are few and
mostly on the synchronicity of price movements across stocks
\cite{mor00,dur04,cha06,wil07}. The Chinese stock market is an
important emerging market. Although sharing certain common features
with developed markets, it also exhibits unusual dynamic properties
\cite{ren05,qiu06,qiu07}. Especially, the anti-leverage effect in
the Chinese market is unique according to up-to-date literatures
\cite{qiu06,qiu07}. In some sense, the Chinese market is a
representative emerging market, and the Indian market is gradually
approaching a developed one.

In this paper, we continue our comparative study of the Chinese
market and developed markets as well as even the Indian market, with
the RMT theory and using the data of the Shanghai Stock Exchange
(SSE) of China. The Shanghai Stock Exchange was established in 1990,
and it is now one of the most important emerging stock markets in
the world. We have analyzed the daily closing prices of 259 stocks
traded in the SSE from Jan., 1997 to Nov., 2007, and it corresponds
to 2633 days. The data are obtained from "Wind Financial Database"
(http://www.wind.com.cn). If the price of a stock is missing on a
particular day, it is then assumed that the price remains the same
as the preceding day \cite{wil07}. In Ref. \cite{pan07a}, it has
been testified that the missing prices do not result in artifacts
for the American market.

Let $P_{i}(t)$ be the price of a stock $i=1 , \ldots , N$ at time t,
we then define the logarithmic price return of the $ith$ stock over
a time interval $\Delta{t}$ as
\begin{equation}
R_{i}(t,\Delta{t})\equiv{\ln{P_{i}(t+\Delta t)}-\ln{P_{i}(t)}}.
\label{e1}
\end{equation}
Since different stocks may fluctuate at different levels, e.g.,
measured by the standard deviation of its return, we should make
sure that the results are independent of the scale of the
measurement. Therefore, we introduce the normalized return
\begin{equation}
r_{i}(t,\Delta t)\equiv{\frac {R_{i}-<R_{i}>} {\sigma_{i}}},
\label{e2}
\end{equation}
where $\sigma_{i}\equiv{\sqrt{<R_{i}^{2}>-<R_{i}>^{2}}}$ is the
standard deviation of $R_{i}$, and $<\ldots>$ represents the time
average. Finally, we compute the equal-time cross-correlation matrix
$\bf C$, whose element
\begin{equation}
C_{ij}\equiv{<r_{i}r_{j}>} \label{e3}
\end{equation}
measures the correlation between the returns of stocks $i$ and $j$.
By definition, $\bf C$ is a real symmetric matrix with $C_{ii}=1$,
and $C_{ij}$ is valued in the domain $[-1,1]$.

\begin{figure}
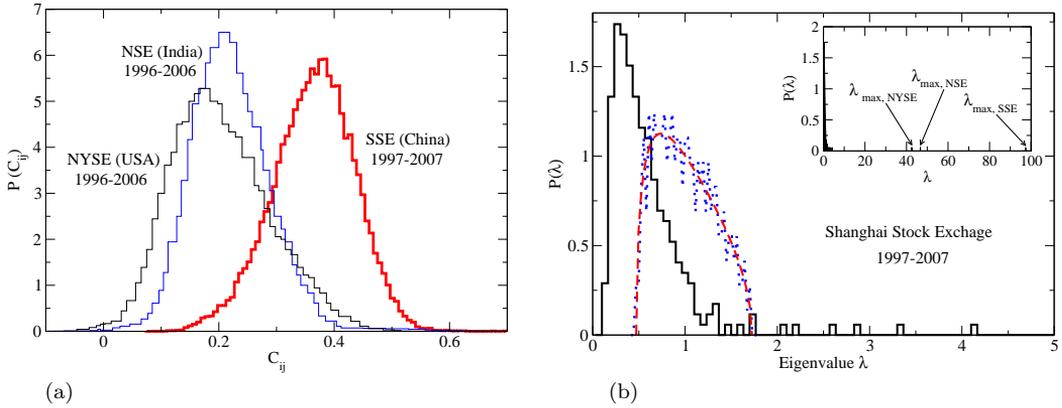

\twoimages[scale=0.28]{CElementPDF.eps}{CEigPDF.eps}
\hspace{0.5cm}\footnotesize{(a)}\hspace{7.0cm}\footnotesize{(b)}
 \caption{(a) The probability distributions of the elements of the
cross-correlation matrix $\bf C$ for the SSE (China), NSE (India)
and NYSE (USA). (b) The probability distribution of the eigenvalues
of the correlation matrix $\bf C$ for the SSE is displayed with the
solid line. The theoretical distribution of the Wishart matrix is
shown with the dashed line, which fits to that of the surrogate
correlation matrix. The inset shows the largest eigenvalue for the
SSE (China), NSE (India) and NYSE (USA), which is 97.33, 46.67 and
42.50 respectively.} \label{f1}
\end{figure}

In Fig. \ref {f1} (a), the probability distribution $P(C_{ij})$ of
the elements of the cross-correlation matrix $\bf C$ is displayed
for the SSE (China), NSE (India) and NYSE (USA). The data for the
NSE and NYSE are from the period 1996-2006 \cite{pan07a}. The mean
value $\left<C_{ij}\right>$ of the elements for the SSE is 0.37,
much larger than 0.22 and 0.20 for the NSE and NYSE respectively
\cite{pan07a}. This result strongly supports the general belief that
stock prices in emerging markets tend to be more correlated than
developed ones \cite{mor00}. In this sense, however, the Chinese
market is much more 'emerging' than the Indian market. In addition,
all elements of the matrix $\bf C$ of the SSE are positive, while
there are a few negative ones for the NSE and NYSE. This also
suggests that the correlation between stocks is stronger in the
Chinese stock market.

If N time series of length $T$ are mutually uncorrelated, the
resulting cross-correlation matrix is called a {\it Wishart} matrix.
Statistical properties of such random matrices are known
\cite{dys71,sen99}. In the limit $N\rightarrow\infty$ and
$T\rightarrow\infty$ with $Q\equiv{T/N}\geq{1}$, the probability
distribution $P_{rm}(\lambda)$ of the eigenvalue $\lambda$ is given
by
\begin{equation}
P_{rm}(\lambda)=\frac{Q} {2\pi} \frac
{\sqrt{(\lambda_{max,ran}-\lambda)(\lambda-\lambda_{min,ran})}}{\lambda}
\label{e4}
\end{equation}
for $\lambda_{min,ran}\leq\lambda\leq\lambda_{max,ran}$ and 0
otherwise. The lower (upper) bound is given by
\begin{equation}
\lambda_{min(max),ran}=[1\pm(1/\sqrt{Q})]^{2}.
\end{equation}
For the data of the SSE in the period 1997-2007, for example, there
are $N=259$ stocks each containing $T=2632$ returns. If there are
{\it no} correlations between the stocks, the eigenvalues should be
bounded between $\lambda_{min,ran}=0.47$ and
$\lambda_{max,ran}=1.73$.

In developed stock markets such as those of USA and Japan, the bulk
of the eigenvalue spectrum $P(\lambda)$ of the cross-correlation
matrix is similar to $P_{rm}(\lambda)$ of the Wishart matrix, but
some large eigenvalues deviate significantly from the upper bound
$\lambda_{max,ran}$ \cite{lal99,ple99,ple02a,uts04}. This is also
true for emerging stock markets, but the number of the large
eigenvalues is relatively few as it is reported for the NSE (India)
\cite{pan07a}. The latter should be a characteristic of emerging
markets. In Fig. \ref {f1} (b), the eigenvalue spectrum of the SSE
(China) is displayed with the solid line. The bulk is similar to
$P_{rm}(\lambda)$ of the Wishart matrix, and there are a few large
eigenvalues. The number of the large eigenvalues is less than that
of the NYSE (USA). To verify that these outliers are not an artifact
of the finite length of the observation period, we shuffle the
empirical time series of returns (i.e., randomly change the time
sequence of returns), thereby destroying all the equal-time
correlations, then compute a surrogate correlation matrix. The
eigenvalue spectrum of this surrogate matrix matches exactly with
$P_{rm}(\lambda)$. This is shown with the dotted line in Fig. \ref
{f1} (b). Therefore, one confirms that the large eigenvalues are the
genuine effect of equal-time correlations between stocks.

The presence of a well-defined bulk of the eigenvalue spectrum which
agrees with $P_{rm}(\lambda)$ suggests that the contents of $\bf C$
are mostly random except for the large eigenvalues that deviate. As
shown in Table \ref{t1} and Fig. \ref {f1} (b), the largest
eigenvalue $\lambda_{0}$ of the cross-correlation matrix of the SSE
(China) is 97.33, about 56 times as large as the upper bound
$\lambda_{max,ran}$ of $P_{rm}(\lambda)$. However, $\lambda_{0}$ of
the NSE (Indian) and NYSE (USA) is 46.67 and 42.50, only about 26
and 28 times as large as $\lambda_{max,ran}$ respectively. The
correlation corresponding to the largest eigenvalue $\lambda_{0}$ is
believed to be generated by interactions common for stocks in the
entire market. Our results imply that this correlation of the market
mode in the Chinese stock market is much stronger than that in other
stock markets, including the NSE (Indian) which is usually
considered to be an emerging market.

\begin{figure}
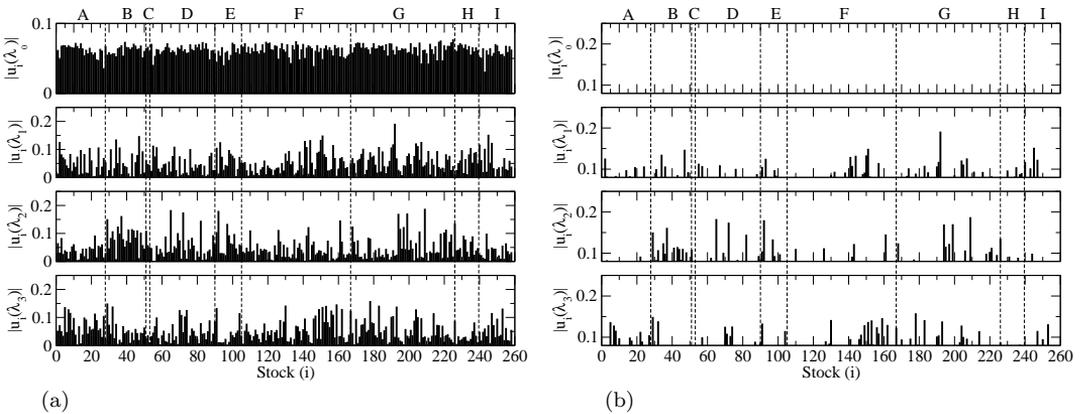

\twoimages[scale=0.28]{CVComponent.eps}{CVComponent008.eps}
\hspace{0.5cm}\footnotesize{(a)}\hspace{7.0cm}\footnotesize{(b)}
\caption{(a) The absolute values of the components $u_{i}(\lambda)$
of stock i corresponding to the first four largest eigenvalues of
the cross-correlation matrix $\bf C$. Results are obtained with the
data of the $SSE$ (China) in the period 1997-2007. Stocks are
arranged according to business sectors separated by dashed lines. A:
Finance; B: IT; C: Energy Resources; D: Basic Materials; E: Daily
Consumer Goods; F: Non-Daily Consumer Goods; G: Industry; H: Public
Utility; I: Medical and Health Care. (b) The same figure as (a), but
only those data with $|u_{i}(\lambda)|\ge u_c=0.08$ are shown.}
\label{f2}
\end{figure}

Now let us look at the components $u_i(\lambda)$ of the eigenvectors
of the first four largest eigenvalues. In the top sector of Fig.
\ref{f2} (a), the eigenvector of the largest eigenvalue
$\lambda_{0}$ shows a relatively uniform composition with all stocks
participating in it. Actually, all the elements also have the same
sign. It represents a common mechanism that affects all the stocks
with the same bias. So the largest eigenvalue is associated with the
market mode, i.e., the collective response of the entire market to
the external information. The Chinese market shares this feature
with other stock markets \cite{pan07a,lal99,ple02a}.

According to previous studies \cite{ple02a,gop01}, the components of
the eigenvectors of other large eigenvalues are localized, i.e.,
each of these eigenvectors is dominated by only a part of stocks.
The dominating components of an eigenvector belong to similar or
related business sectors. This is true for both developed markets
and some emerging markets. However, the number of the large
eigenvalues of the NYSE (USA) is obviously more than those of the
NSE (India) \cite{gop01,ple02a,pan07a}. That of the SSE (China) is
even less. Especially, the dynamic effect of business sectors is
hardly seen, as shown in the second, third and fourth sectors of
Fig. \ref{f2} (a).

{\it What interactions do control the large eigenvalues of the SSE?}
To answer it, we introduce a threshold $u_c$ to select the
dominating components of the eigenvectors with large eigenvalues. In
other words, we consider that only those components which satisfy
$|u_i(\lambda)|\ge u_c$ contribute to the eigenvector of the
eigenvalue $\lambda$. The results of $u_c=0.08$ is shown in Fig.
\ref{f2} (b).

In the SSE (China), a company will be specially treated when its
financial situation is abnormal. In this case, the word "ST" will
then be added to the stock name as a prefix. The abnormal financial
situation includes: the audited profits are negative in two
successive accounting years, the audited net worth per share is less
than its stock's par value in the recent accounting year etc. When
the company has a negative audited profit for three successive
accounting years, the word "*ST" will be added to its stock name and
the stock may be also delisted. When the abnormal financial
conditions are eliminated, the word "ST" or "*ST" will be removed.

For the SSE (China), we observe that the eigenvector of the second
largest eigenvalue $\lambda_{1}$ is dominated by the stocks now with
or used to be with "ST" or "*ST". When we increase the threshold
$u_c$, the proportion of the dominating stocks with "ST" and "*ST"
rises. When $u_c\ge 0.12$, all dominating components are "ST" or
"*ST" stocks. Detailed results are given in Table \ref{t2}.
Therefore, the second largest eigenvalue corresponds to the
so-called ST sector.

The dominating components of the eigenvector of the third largest
eigenvalue $\lambda_{2}$ mainly are or used to be "Blue-chip"
stocks, which are referred to be with stable and good performance,
i.e., a reasonable positive profit in a period of time. When we
increase the threshold $u_c$, the proportion of the Blue-chip stocks
steadily rises, as shown in Table \ref{t2}. With a similar approach,
we examine the dominating components of the eigenvector of the forth
largest eigenvalue $\lambda_{3}$, and observe that they are mainly
those companies registered in Shanghai with the real estate
business. When we continue to investigate the next largest
eigenvalues such as $\lambda_{4}$ and $\lambda_{5}$ of the SSE
(China), we could not find clear characteristics. For the NYSE
(USA), however, one could detect the dominating effect of the
business sectors at least up to the 10th largest eigenvalue
\cite{gop01,ple02a}.

{\it Why does the SSE (China) exhibit the usual sectors like the ST
and Blue-chip sectors rather than the standard business sectors?}
Since the Chinese stock market is an emerging market, the companies
are not operated strictly with the registered business. On the other
hand, people seriously look at the performance of the companies. The
fourth largest eigenvalue of the SSE may reflect the fact that
Shanghai and especially its real estate business play important
roles in China in the past years both economically and politically.

To better understand the structure of interactions in the SSE
(China), we introduce a variation of the two-factor model of the
market dynamics \cite{pan07a}. In our model, we assume that the
normalized return at time t of the $i$th stock from the $k$th
business sector can be decomposed into (i) a market factor
$r_{m}(t)$, that contains information common to all stocks; (ii) a
business sector factor $r_{g}^{k}(t)$, representing dynamic effects
exclusive to stocks in the $k$th business sector; (iii) a profit
factor $r_{p}(t)$, reflecting three categories of stocks: the "ST"
sector, "Blue-chip" sector and "general" sector--i.e., the rest
stocks except the ST and Blue-Chip stocks; and (iv) an idiosyncratic
term $\eta_{i}(t)$, which corresponds to random variations unique
for that stock. We thus obtain
\begin{equation}
r_{i}^{k}(t)=\beta_{i}r_{m}(t)+\gamma_{i}^{k}r_{g}^{k}(t)+
\gamma_{i}^{p}r_{p}(t)+\sigma_{i}\eta_{i}(t) \label{e7}
\end{equation}
where the constants $\beta_{i}$, $\gamma_{i}^{k}$, $\gamma_{i}^{p}$
and $\sigma_{i}$ represent the relative strengths of the four
factors respectively. Although the empirically distributions of
returns show power-law tails and are not Levy stable, we assume that
the normalized returns $r^{k}_{i}$ follow a Gaussian distribution
with a zero mean and a unit variance for simplicity. In other words,
we choose $r_{m}(t)$, $r_{g}^{k}(t)$, $r_{p}(t)$ and $\eta_{i}(t)$
to be Gaussian processes with a zero mean and a unit variance. The
unit variance ensures that the relative strengths of the four
factors satisfy the relation
\begin{equation}
{\beta_{i}}^{2}+{(\gamma_{i}^{k})}^{2}+{(\gamma_{i}^{p})}^{2}+{\sigma_{i}}^{2}=1.
\label{e8}
\end{equation}
For each stock, we can independently assign $\gamma_{i}^{k}$,
$\sigma_{i}$ and $\gamma_{i}^{p}$, and obtain $\beta_{i}$ from Eq.
(\ref {e8}). But we should assure that all $\gamma_{i}^{p}$ for the
ST sector are greater than any $\gamma_{i}^{p}$ for the Blue-chip
sector. This condition may lead to the eigenvalue spectrum of the
SSE (China), whose second and third largest eigenvalues represents
the ST and Blue-chip sectors respectively. We practically choose
$\gamma_{i}^{k}$, $\sigma_{i}$ and $\gamma_{i}^{p}$ from a uniform
distribution with a width $\delta$ and centered about the mean value
$\gamma$, $\sigma$ and $\gamma^{p}$ respectively.

With the above model, we generate $N$ times series of length $T$ for
returns $r_{i}^{k}$, and simulate an artificial market with $N$
stocks and $K$ business sectors. These $K$ business sectors are
composed of $n_{1}$, $n_{2}$, $\ldots$, $n_{k}$ stocks such that
$n_{1}+n_{2}+\ldots+n_{k}=N$. Next we randomly select $n_{ST}$
stocks as "ST" stocks and $n_{BC}$ stocks as "Blue-Chip" stocks from
the N stocks such that $n_{ST}+n_{BC}+n_{G}=N$, where $n_{G}$
represents the number of the rest stocks after selecting ST and
Blue-chip stocks. Finally, we construct the resultant
cross-correlation matrix $\bf C$ and compute its eigenvalues and
eigenvectors.

In order to describe the empirical properties of the SSE (China), we
choose a large $\beta$, moderate $\gamma^{p}$ and small $\gamma$.
This leads to a market with a strong market mode, moderate
interactions between stocks in the ST or Blue-chip sectors, and a
small effect of the business sectors. For example, we take $N=250$
and $T=2500$, and assign 5 sectors A, B, C, D and E, each with 50
stocks. We choose $\gamma_{i}^{k}=0.2$, $\gamma_{i}^{p}=0.55$ for
the ST sector and $\gamma_{i}^{p}=0.40$ for the Blue-Chip sector,
and $\sigma=0.3$, $\delta=0.05$, $n_{ST}=40$ and $n_{BC}$=40. With
these fixed parameters, one calculates $\beta$ to be 0.75 and 0.84
for the ST and Blue-chip sectors respectively.

\begin{figure}
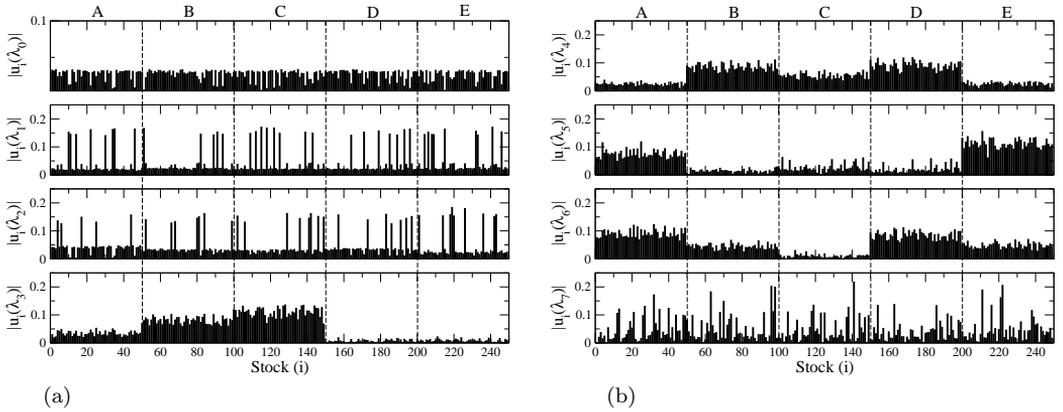

\twoimages[scale=0.28]{Model3CVComponent1.eps}{Model3CVComponent2.eps}
\hspace{0.5cm}\footnotesize{(a)}\hspace{7.0cm}\footnotesize{(b)}
\caption{(a) The absolute values of the components $u_{i}(\lambda)$
of stock i corresponding to the first four largest eigenvalues of
the cross-correlation matrix $\bf C$ from numerical simulations. The
stocks are separated by dashed lines according to business sectors.
(b) A similar figure as (a), but for next four largest eigenvalues
of $\bf C$ from numerical simulations. $\lambda_{7}$ corresponds to
the upper bound $\lambda_{max,ran}$ of the Wishart matrix. }
\label{f4}
\end{figure}

In Fig. \ref {f4}, the components of the eigenvectors of the first
eight largest eigenvectors are displayed. The largest eigenvalue
represents the market mode, and the components are distributed with
uniform weights. For the second and third eigenvalues, we observe
large components from different business sectors. When we introduce
a threshold $u_c$ to select those dominating components which
satisfy $|u_i| \ge u_c$, we find that they are all ST stocks for the
second largest eigenvalue and Blue-chip stocks for the third largest
eigenvalue.

For the next largest eigenvalues, we detect the dynamic effect of
the standard business sectors. But several business sectors may
contribute to a single eigenvector. This behavior is also similar to
that of the Chinese stock market. The eighth largest eigenvalue in
Fig. \ref {f4} is the upper bound $\lambda_{max,ran}$ of the Wishart
matrix which is derived from uncorrelated time series. In our
numerical simulations, the cross-correlation matrix is essentially
modified by the extra interactions between the stocks in the unusual
sectors such as the ST and Blue-chip sectors.

In summary, we phenomenologically compute the correlations of stock
price returns, and construct the cross-correlation matrix $\bf C$.
The average value $\left<C_{ij}\right>$ of the matrix elements is
0.37 for the SSE (China), much larger than 0.22 and 0.20 for the NSE
(Indian) and NYSE (USA) respectively. It supports that the
correlations in emerging markets are stronger than those in
developed markets. In this sense, the NSE is obviously less emerging
than the SSE.

We then diagonalize the matrix $\bf C$ and obtain the eigenvalue
spectrum and eigenvectors. The large eigenvalues which go beyond the
upper bound $\lambda_{max,ran}$ of the Wishart matrix reflect the
correlations between stocks. The largest eigenvalue $\lambda_0$
corresponds to the market mode, which is induced by interactions
common for all stocks in the stock market. The largest eigenvalue
$\lambda_{0}$ of the SSE (China) is 97.33, much large than 46.67 of
the NSE (Indian) and 42.50 of the NYSE (USA) respectively. This
result is in agreement with that of the average value
$\left<C_{ij}\right>$. Other large eigenvalues represent the
interactions between stocks in a same sector. For the NYSE, the
correlations corresponding to standard {\it business} sectors are
strong, and there are at least 10 such large eigenvalues. For
emerging markets such as the SSE and NSE, the correlations between
stocks in a same sector are weak, and there exist only a few large
eigenvalues. Especially, the correlations within a standard business
sector is almost invisible. For the SSE (China), we identify the
large eigenvalues according to unusual sectors such as the ST and
Blue-chip sectors.

Finally we apply a variation of the two-factor dynamic model to
explain the collective behavior of the emerging stock markets. In
our model, extra interactions within an unusual sector are
introduced, and the correlations corresponding to the standard
business sectors are then suppressed.

{\bf Acknowledgements:} This work was supported in part by NNSF
(China) under grant No. 10875102.


\newpage

\begin{table}[p]\centering
\caption{The values of $T, N, Q,$ $\lambda_{min(max), ran}$ and
$\lambda_{min(max), real}$ of the SSE (China), in comparison with
those of the NSE (India) and NYSE (USA) \cite{pan07a}.
$\lambda_{min(max), ran}$ represents the low (upper) bound of the
eigenvalues of the Wishart matrix, while $\lambda_{min(max), real}$
are those of real stock markets.}
 \small\begin{tabular}{c | c c c c c c c c c}
     & Period  & T   & N & Q   & $\lambda_{min,
ran}$ & $\lambda_{max, ran}$
 & $\lambda_{min, real}$ & $\lambda_{max, real}$\\
\hline
SSE (China)  & 1997-2007 & 2632  & 259 & 10.16 & 0.47 & 1.73 & 0.18  & 97.33 \\
NSE (India)  & 1996-2006 & 2621  & 201 & 13.04 & 0.52 & 1.63 & 0.20  & 46.67 \\
NYSE (USA)   & 1996-2006 & 2606  & 201 & 12.97 & 0.52 & 1.63 & 0.31  & 42.50 \\

\hline\end{tabular} \label{t1}
\end {table}

\begin{table}[p]\centering
\caption{The proportions of "ST", "Blue chip" and "Shanghai Real
Estate" (SHRE) stocks for the eigenvectors of the second, third and
fourth largest eigenvalues respectively. Results are given for two
thresholds $u_c$. The unit is percentage.}
\begin{tabular}{l| l| l| l }
       & $u_i(\lambda_1)$  & $u_i(\lambda_2)$    & $u_i(\lambda_3)$   \\
\hline
$u_c$  & ST  & Blue-chip    & SHRE  \\
\hline
0.08          & 80.00 & 73.91  & 81.48 \\
0.12          & 100.00 & 83.33  & 75.00 \\
\hline

\end{tabular}
\label{t2}
\end {table}

\end{document}